\documentclass[10pt,journal,twocolumn,twoside]{IEEEtran}
\usepackage{amsfonts,color,xcolor}
\usepackage{amssymb,amsmath,mathrsfs}
\usepackage{bm}
\usepackage{mathtools}
\mathtoolsset{showonlyrefs}
\usepackage[noadjust]{cite}
\usepackage[pdftex]{graphicx}
\usepackage{epstopdf}
\usepackage{float}
\usepackage{relsize}

\definecolor{amGreen}{HTML}{6B8E23}

\definecolor{IndianRed3}{rgb}{0.8,0.33,0.33}

\title{Information Freshness Analysis of Slotted ALOHA in Gilbert-Elliot Channels}
\author{Andrea Munari, \IEEEmembership{Senior Member, IEEE}, and Gianluigi Liva, \IEEEmembership{Senior Member, IEEE}
\thanks{A. Munari and G. Liva are with the Institute of Communications and Navigation of the German Aerospace Center (DLR), 82234 Wessling, Germany. Email:\{\texttt{Andrea.Munari}, \texttt{Gianluigi.Liva}\}\texttt{@dlr.de}.}
\vspace{-1em}}

\newtheorem{remark}{Remark}
\newtheorem{example}{Example}

\newcommand{\de}{\mathrm{d}}

\newcommand{\pr}{\ensuremath{\mathbb P}}
\newcommand{\expOp}{\ensuremath{\mathbb E}}

\newcommand{\nodes}{\ensuremath{n}}

\newcommand{\ps}{\ensuremath{\mathrm p_{s}}}

\newcommand{\pGen}{\ensuremath{\alpha}}
\newcommand{\pGB}{\ensuremath{\beta}}
\newcommand{\pBG}{\ensuremath{\gamma}}
\newcommand{\pStatCh}{\ensuremath{\pi}}
\newcommand{\good}{\ensuremath{\mathsf G}}

\newcommand{\pGood}{\ensuremath{\pStatCh_{\good}}}

\newcommand{\lastUpd}{\ensuremath{\tau}}
\newcommand{\penFun}{\ensuremath{\delta}}
\newcommand{\penFunM}{\ensuremath{\penFun^{(m)}}}
\newcommand{\penFunMRP}{\ensuremath{\Delta^{(m)}}}
\newcommand{\avPenFun}{\ensuremath{\bar{\Delta}}}
\newcommand{\avPenFunM}{\ensuremath{\avPenFun^{(m)}}}
\newcommand{\avPenFunTwo}{\ensuremath{\avPenFun^{(2)}}}
\newcommand{\peakPen}{\ensuremath{\Xi}}
\newcommand{\peakPenM}{\ensuremath{\peakPen^{(m)}}}

\newcommand{\UD}{\ensuremath{Y}}
\newcommand{\ud}{\ensuremath{y}}

\renewcommand{\succ}{\ensuremath{\mathsf S}}
\newcommand{\miss}{\ensuremath{\mathsf M}}
\newcommand{\block}{\ensuremath{\mathsf B}}

\usepackage{acro}
\DeclareAcronym{AWGN}{short = AWGN ,long = additive white gaussian noise}
\DeclareAcronym{AoI}{short = AoI ,long = age of information}
\DeclareAcronym{AoII}{short = AoII, long = age of incorrect information}
\DeclareAcronym{PAoI}{short = PAoI ,long = peak age of information}
\DeclareAcronym{CDF}{short = CDF ,long = cumulative distribution function}
\DeclareAcronym{CRA}{short = CRA ,long = contention resolution ALOHA}
\DeclareAcronym{CRDSA}{short = CRDSA ,long = contention resolution diversity slotted ALOHA}
\DeclareAcronym{CP}{short = CP ,long = contention period}
\DeclareAcronym{CSA}{short = CSA ,long = coded slotted ALOHA}
\DeclareAcronym{C-RAN}{short = C-RAN ,long = cloud radio access network}
\DeclareAcronym{DAMA}{short = DAMA ,long = demand assigned multiple access}
\DeclareAcronym{DSA}{short = DSA ,long = diversity slotted ALOHA}
\DeclareAcronym{eMBB}{short = eMBB ,long = enhanced mobile broadband}
\DeclareAcronym{FEC}{short = FEC ,long = forward error correction}
\DeclareAcronym{GEO}{short = GEO ,long = geostationary orbit}
\DeclareAcronym{GF}{short = GF ,long = generating function}
\DeclareAcronym{HAP}{short = HAP ,long = high-altitude platform,foreign-plural={}}
\DeclareAcronym{IC}{short = IC ,long = interference cancellation}
\DeclareAcronym{IoT}{short = IoT ,long = internet of things}
\DeclareAcronym{IRSA}{short = IRSA ,long = irregular repetition slotted ALOHA}
\DeclareAcronym{LEO}{short = LEO ,long = low Earth orbit}
\DeclareAcronym{M2M}{short = M2M ,long = machine-to-machine}
\DeclareAcronym{MAC}{short = MAC ,long = medium access}
\DeclareAcronym{MPR}{short = MPR ,long = multi-packet reception}
\DeclareAcronym{MTC}{short = MTC ,long = machine-type communications}
\DeclareAcronym{mMTC}{short = mMTC ,long = massive machine-type communications}
\DeclareAcronym{MC}{short = MC ,long = Markov chain}
\DeclareAcronym{NTN}{short = NTN ,long = non-terrestrial network,foreign-plural = {}}
\DeclareAcronym{PDF}{short = PDF ,long = probability density function}
\DeclareAcronym{PER}{short = PER ,long = packet error rate}
\DeclareAcronym{PLR}{short = PLR ,long = packet loss rate}
\DeclareAcronym{PMF}{short = PMF ,long = probability mass function}
\DeclareAcronym{RA}{short = RA ,long = random access}
\DeclareAcronym{RRH}{short = RRH ,long = remote radio head,foreign-plural = {}}
\DeclareAcronym{SA}{short = SA , long = slotted ALOHA}
\DeclareAcronym{SFG}{short = SFG , long = signal flow graph}
\DeclareAcronym{SIC}{short = SIC ,long = successive interference cancellation}
\DeclareAcronym{SIR}{short = SIR ,long = signal to interference ratio}
\DeclareAcronym{SNIR}{short = SNIR ,long = signal-to-noise and interference ratio}
\DeclareAcronym{SINR}{short = SINR ,long = signal-to-interference and noise ratio}
\DeclareAcronym{SNR}{short = SNR ,long = signal-to-noise ratio}
\DeclareAcronym{TDM}{short = TDM ,long = time division multiplexing}

\begin{document}

\maketitle
\thispagestyle{empty} \setcounter{page}{0}

\vspace{-1em}
\begin{abstract}
This letter analyzes a class of information freshness metrics for large IoT systems in which terminals employ slotted ALOHA to access a common channel. Considering a Gilbert-Elliot channel model, information  freshness is evaluated through a penalty function that follows a power law of the time elapsed since the last received update, in contrast with the linear growth of age of information. By means of a signal flow graph analysis of Markov processes, we provide exact closed form expressions for the average penalty and for the peak penalty violation probability. 
\end{abstract}

\vspace{-3mm}

\begin{keywords}
Grant-free access, slotted ALOHA, information freshness, age of information, Gilbert-Elliot channels.
\end{keywords}

\section{introduction} \label{sec:intro}

\PARstart{M}{any} key \ac{IoT} applications are characterized by the presence of a large number of terminals that track a process and report time-stamped updates to a sink over a shared wireless channel. This is the case, among others, of industrial automation, fleet and asset tracking, as well as environmental monitoring. In these settings, the ability to maintain an up-to-date view of the observed quantities is paramount, and non-trivial tradeoffs emerge. Indeed, changes in the update transmission frequency directly reflect on the channel contention level, and critically influence the time needed to successfully refresh the monitor's perception. 

These remarks have triggered a florid line of research on \emph{information freshness} \cite{Yates20_Survey}, with the proposal of novel system design criteria and performance metrics. Among these, a precursor role was played by \ac{AoI}, which measures the time elapsed since the generation of the last received status-update. In spite of its simple definition, \ac{AoI} has shed light on some fundamental tradeoffs in point-to-point links (see \cite{Yates20_Survey} for an excellent survey), and, recently, on the more complex behaviour of practical \ac{IoT} settings in which devices share the channel via grant-free policies, e.g. \cite{Yates17:AoI_SA,Munari21_TCOM_AoI,Uysal21_AlohaThresh}. 

On the other hand, the notion of \ac{AoI} presents some intrinsic shortcomings, as its linear growth over time may fail to capture the critical effect of stale knowledge, e.g. when status information is employed in feedback control systems. Furthermore, \ac{AoI} does not allow to model the impact of \emph{incorrect} information available at the receiver end. From this viewpoint, the use of different \emph{penalty functions}, such as the \ac{AoII} \cite{Ephremides20_AoII}, has recently emerged as a fundamental step forward in the study of information freshness. Important results in this direction have been obtained for a single transmitter-receiver setup \cite{Yates20_Survey,Ephremides20_AoII,Uysal20_TIT}. Interestingly, in grant-free multi-user systems studies have only looked at the classical \ac{AoI} metric, to the best of the authors' knowledge.

To tackle this gap, we study a class of information freshness penalty functions that follow a power law of the time elapsed since the last received update. The analysis applies to a setup in which a large number of terminals access the channel via a \ac{SA} policy. Each wireless user-sink link is modelled as a Gilbert-Elliot channel \cite{Gilbert60,Elliott63}, with data losses that are independent across links. This channel model captures in a simple way correlated packet losses that are typical of wireless channels \cite{turin1999digital,Zorzi:GE}, and it is especially relevant to satellite IoT scenarios, where the terminal-satellite links may be intermittently shadowed \cite{lutz2012satellite}. Leaning on a \ac{SFG} \cite{shannon1942theory} analysis of Markov processes, we provide exact closed form expressions for both average penalty and peak penalty violation for any penalty order. The presented framework offers broadly applicable results, providing as special cases an analysis of non-linear freshness metrics for \ac{SA} as well as the first study of the \ac{AoI} of the access scheme under Gilbert-Elliot channels. Non-trivial insights emerge, leading to useful design hints.

\section{System Model and Preliminaries} \label{sec:sysModel}

We focus on a slotted ALOHA system \cite{Abramson77:PacketBroadcasting} without feedback and retransmissions, in which \nodes\ terminals share a common channel to attempt delivery of data towards a common receiver (sink).\footnote{We note that the considered ALOHA system is inherently stable.} Time is divided in slots of equal duration, chosen to allow the transmission of one packet, and all devices are assumed to be slot-synchronous. Without loss of generality, we consider a unit slot duration (i.e., all time-related quantities are expressed in terms of slots). At the beginning of a slot, each node independently becomes active with probability \pGen, generating a status update and sending it over the channel.

The channel between a transmitter and the receiver is described by a Gilbert-Elliot model \cite{Gilbert60,Elliott63}, so that  the link over a slot may be in either a \emph{good} or \emph{bad} state. In the former case, a packet transmitted by the device arrives unfaded at the sink, whereas a data unit is \emph{erased} when the channel is in bad state, bringing no power contribution at the receiver. Transitions in the two-state Markov chain take place at the beginning of each slot, with the channel moving from good to bad with probability $\pGB$ and from bad to good with probability $\pBG$. Accordingly, we denote the stationary probability for the  channel to be in good state as \mbox{$\pGood = \pBG/(\pGB + \pBG)$}. Links from each transmitter to the receiver are assumed to be independent.

Adopting the widely employed collision channel model, we further assume that the presence of two or more unerased packets over a slot (i.e., a collision) prevents decoding of any of them. Conversely, if a single data unit reaches the receiver unfaded, its content is correctly retrieved.

\begin{figure}
  \centering
  \includegraphics[width=.9\columnwidth]{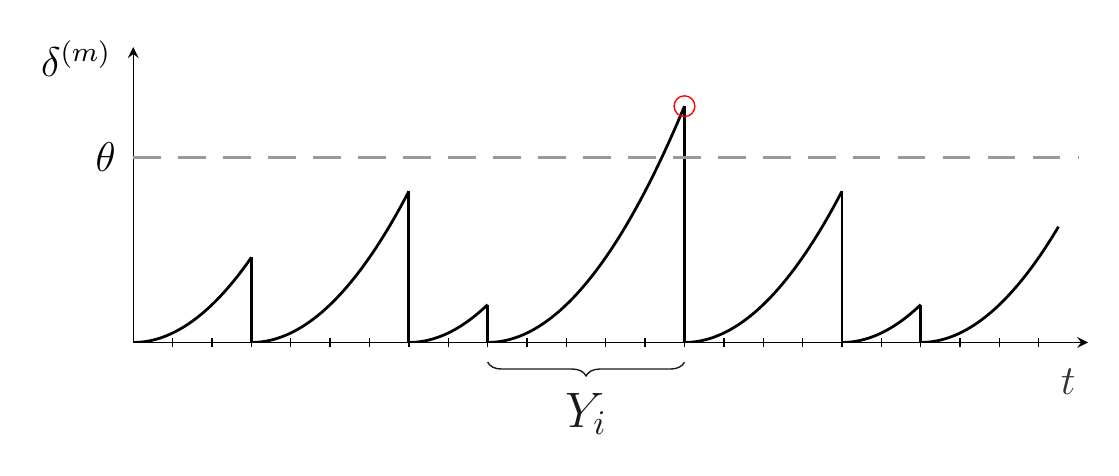}
  \vspace{-1em}
  \caption{Example of time evolution for $\penFunM(t)$, $m=2$. An instance of peak penalty violation is highlighted by the red circle.}
  \label{fig:timelinePenalty}
  \vspace{-.5em}
\end{figure}

In this setup, the sink aims at keeping an up to date knowledge of the status of each node in the network. To gauge this ability, let us focus on a generic source, and denote by $\lastUpd(t)$ the instant at which its latest update was received by the sink, as of time $t$. Leaning on this, we introduce for the source under observation  the \emph{information freshness penalty} function
\begin{equation}
  \penFunM(t) := (t - \lastUpd(t))^m
  \label{eq:def_penFun}
\end{equation}
with $m\in\mathbb N^+$. Note that, for $m=1$, \eqref{eq:def_penFun} returns the well-known \emph{age of information} metric \cite{Yates20_Survey}. The definition introduces however a more general notion of penalty, and provides for $m>1$ a flexible model to describe the stronger repercussions that a stale knowledge may have.
We denote the corresponding random process as $\penFunMRP(t)$. An example of the evolution of $\penFunM(t)$ is reported in Fig.~\ref{fig:timelinePenalty}, for the case $m=2$. As illustrated in the plot, each time an update from the node is received, the penalty function is reset to $0$, since the sink gathers the freshest possible knowledge of the status of the source. Conversely, as time elapses since the last collected update, $\penFunM(t)$ increases following the power law \eqref{eq:def_penFun}, and captures the impact of having outdated information on the monitored device.

As will be discussed in Sec.~\ref{sec:analysis}, the process $\penFunMRP(t)$ is ergodic, and we will focus in the remainder of our discussion on its stationary behaviour. Specifically, we consider two quantities to evaluate the system performance:  \emph{average penalty} and \emph{peak violation probability}. The former is introduced as
\begin{equation}
  \avPenFunM := \expOp\left[ \,\penFunMRP(t)\,\right].
\label{eq:avPenFun_def}
\end{equation}
The latter, instead, is denoted as $\peakPenM(\theta)$ and defined as the probability that the maximum value of the penalty function within one update cycle, i.e. the value of $\penFunM$ right before an update is received, exceeds a threshold $\theta$ (see also Fig.~\ref{fig:timelinePenalty}). The metric offers an important design tool in settings where relying on exceedingly stale information may not be tolerated (e.g., control systems), and captures in the special case $m=1$ the peak age violation probability \cite{Chiariotti21_peakAge,Durisi19_JSAC}. We furthermore note that the presented analytical approach can be extended to a wider class of penalty functions \cite{NLAoI} with minor effort, as will be discussed at the end of Sec.~\ref{sec:analysis}.

\textbf{Notation:} Throughout our discussion, we denote a r.v. (or random process) and its realizations with capital and small letters, respectively. For a non-negative discrete r.v. $A$ taking values in $\mathbb N^0$, we indicate its \ac{PMF} as $P_A(a)$, and refer to its \ac{GF}
\begin{align}
  G_A(x) := \sum_{a=0}^\infty P_A(a)\, x^a = \expOp\left[ x^A\right].
  \label{eq:genFun_def}
\end{align}
We denote the corresponding \emph{moment} generating function as $M_A(s)$, obtained by setting $x=e^s$ in \eqref{eq:genFun_def}. 
Leaning on $M_A(s)$, the $k$-th order moment of $A$ can readily be derived as 
\begin{align}
  \expOp\left[A^k\right] = \left.\frac{\de^k \!M_A(s)}{\de s^k} \right|_{{\scriptstyle s=0}}.
\label{eq:momentK_genFun}
\end{align}
Finally, we denote by $q_{(i,j)}$ the one-step transition probability between states $i$ and $j$ of a discrete time \ac{MC}.

\section{Analysis} \label{sec:analysis}

The evolution of the penalty function $\penFunMRP(t)$ may be completely specified by studying the embedded discrete Markov process $(C_\ell,\penFunMRP_\ell)$, which captures the state of the channel $(C_\ell)$ and the value of the penalty $(\penFunMRP_\ell)$ at the start of the $\ell$-th slot. Albeit conceptually viable, however, this task quickly becomes cumbersome in the considered setting.

We follow thus a different approach, and focus on the discrete process $\UD_i$, tracking the duration of the $i$-th update cycle, i.e. the number of slots elapsed between the reception of the $(i\!-\!1)$-th and the $i$-th update from the source of interest (see Fig.~\ref{fig:timelinePenalty}). As a first remark, we note that the process is i.i.d. and thus ergodic. This readily follows from the independent behaviour of the source across slots, and by observing how each cycle initiates in the same conditions, starting right after the reception of an update which requires the channel to be in good state. We denote the \ac{PMF} of such process as $P_\UD(\ud)$.

More interestingly, the statistical characterization of $\UD$ suffices to compute the performance metrics of interest. Consider first the average penalty defined in \eqref{eq:avPenFun_def}. The quantity can be conveniently derived conditioning on the duration $Y(t)$ of the update cycle over which the observation time $t$ lies, to obtain
\begin{align}
  \avPenFunM \!=\! \sum\nolimits_{y} \expOp\left[ \, \penFunMRP(t) \,\middle |\, \UD(t) = y \right] \pr\{ \UD(t) = y\} \label{eq:avPenFun_deriv_step1}
\end{align}
where the probability for $t$ to fall within an update cycle of duration $\UD(t)=\ud$ can promptly be expressed as
\begin{align}
  \pr\{\UD(t) = y\} = \frac{\ud \, P_\UD(\ud)}{\sum\nolimits_{\ud'} \ud' \,P_\UD(\ud')} \,.
\end{align}
In turn, the observation time follows a uniform distribution conditioned on $Y(t)$. Therefore, recalling \eqref{eq:def_penFun}  we have
\begin{align}
  \!\!\!\avPenFunM  \!=\! \sum\nolimits_\ud \int_{0}^{\ud} \!\frac{u^m}{\ud} \,\de u \cdot  \frac{\ud \, P_\UD(\ud)}{\sum\nolimits_{\ud'} \ud' \,P_\UD(\ud')} \!=\!
\frac{\expOp \!\left[ \UD^{m+1}\right]}{(m+1) \,\expOp\left[\UD\right]} \,\,\,
\label{eq:avPenFun_moments}
\end{align}
which shows how the average penalty can be derived for any value of $m$ once the moments of $\UD$ are known.

\begin{figure}
  \centering
  \includegraphics[width=.72\columnwidth]{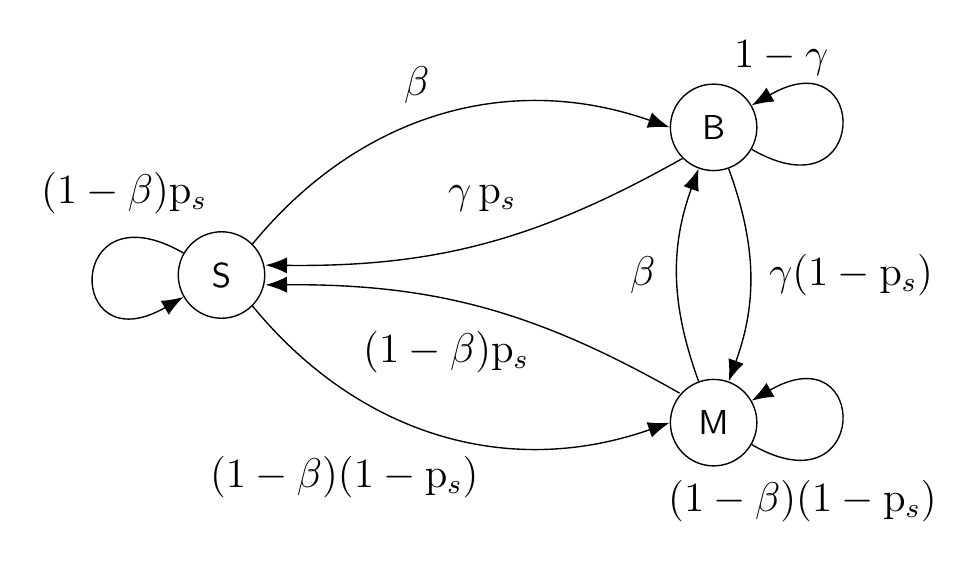}\vspace{-3mm}
  \caption{Markov chain tracking the evolution of the inter-update time $\UD$.}\vspace{-3mm}
  \label{fig:MC_Y}
\end{figure}

Similarly, the peak violation probability can be readily characterised observing how the maximum value taken by the penalty metric within an update interval of duration $\UD=\ud$ slots is exactly $\ud^m$, so that
\begin{align}
  \peakPenM(\theta) = \pr\{\UD^m > \theta \}.
  \label{eq:peakPen_momnets}
\end{align}
Leaning on \eqref{eq:avPenFun_moments} and \eqref{eq:peakPen_momnets} we thus study the stochastic process $\UD_i$. To this aim, consider the discrete time \ac{MC} reported in Fig.~\ref{fig:MC_Y}. Here, three states are identified: \succ, denoting a slot over which an update from the source is successfully received; \miss, indicating a slot where the channel was in good state but no update was delivered (either due to a collision or to the lack of transmission by the source); and \block, denoting a slot with bad channel conditions for the source. To specify the transition probabilities of the chain, let us furthermore introduce the quantity \ps, denoting the probability for a node with \emph{good} channel to the sink to deliver an update. For the system model under study, this evaluates in stationary conditions to
\begin{align}
  \ps = \pGen (1-\pGen \pGood)^{\nodes-1}
  \label{eq:psucc}
\end{align}
where the first factor accounts for the node to actually generate and transmit a packet, whereas the latter for  no other unerased packet to reach the sink (i.e., absence of collision). With this notation, let us then focus on state \succ, and recall that the delivery of an update implies that the source-sink channel was in good state over the slot. Accordingly, the chain transitions in the next slot to \block\ if the channel becomes bad, irrespective of the nodes' activity, i.e., with probability $q_{(\succ,\block)} = \pGB$. On the other hand, the chain moves to \miss\ if the channel remains good, yet the source does not deliver an update, i.e., with probability $q_{(\succ,\miss)} = (1-\pGB) (1-\ps)$. The process remains in state \succ\ if another update can be delivered in the subsequent slot, i.e. with probability $q_{(\succ,\succ)} = (1-\pGB) \ps$. Following a similar reasoning all other transition probabilities are reported in Fig.~\ref{fig:MC_Y}.

The discussed \ac{MC} allows to characterize the duration of an update cycle, as the process $\UD_i$ is exactly captured by the recurrence time of state \succ\ (i.e., the number of slots required to return to \succ\ when starting from there). To derive the corresponding statistics, we resort to a \ac{SFG} analysis.\footnote{An interesting \ac{SFG}-based approach to study linear penalty functions, i.e., \ac{AoI}, for point-to-point M/G/1/1 preemptive queues can also be found in \cite{Telatar18}.}

\subsection{Signal Flow Graph Analysis}
A \ac{SFG} consists of a set of nodes connected by directed weighted edges. Each node $v$ is associated to a variable, whose value is the weighted sum of the variables of nodes with a directed edge entering $v$. In other words, \ac{SFG}s offer a graphical representation of systems of linear equations. Based on this parallel, a broad set of techniques have been proposed for their study \cite{Lorens56}. Among these, Mason's gain formula provides a simple algorithm to compute the direct dependency between two nodes' variables \--- the so-called transfer function \---  through  a visual inspection of the graph \cite{Mason55}.
For Markov processes, SFGs are an effective tool to capture recurrence times. Specifically, consider the \ac{SFG} whose nodes and edges correspond to the transition diagram of a \ac{MC}, and in which the weight of an edge $(i,j)$ is set as $x\cdot q_{(i,j)}$, where $x$ is a dummy variable. In this case, the transfer function between a generic node $v$ and an absorbing state (i.e., a node with no outgoing edges) corresponds to the generating function of the absorption time when starting from $v$ \cite{Lorens56}. 

\begin{figure}
  \centering
  \includegraphics[width=.62\columnwidth]{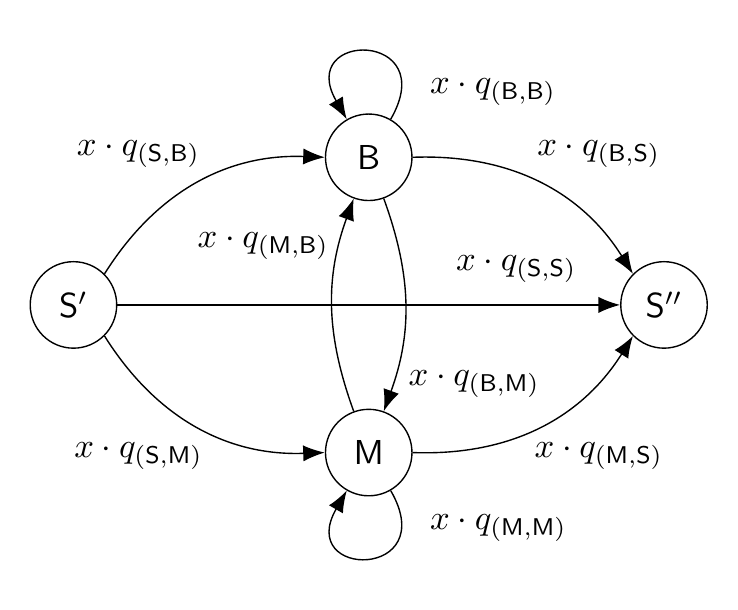}
  \vspace{-1.2em}
  \caption{Signal flow graph derived from the \ac{MC} of Fig.~\ref{fig:MC_Y}, which allows to derive the generating function of \UD\ using Mason's gain formula.}\vspace{-2mm}
  \label{fig:SFG_Y}
\end{figure}

Fig.~\ref{fig:SFG_Y} shows how this approach can be applied to the system under study. Starting from the \ac{MC} of Fig.~\ref{fig:MC_Y}, state \succ\ is split into $\succ^\prime$ (emanating transitions out of state \succ) and $\succ^{\prime\prime}$ (collecting all transitions that lead to a success). In this way, the absorption time from $\succ^\prime$ into $\succ^{\prime\prime}$ captures exactly the r.v. \UD, i.e. the time between two successive update deliveries. 
To complete the construction of the \ac{SFG}, edge weights are set as discussed based on the one-step transition probabilities of the original MC. The generating function $G_{\UD}(x)$ is then given by the transfer function between $\succ^\prime$ and $\succ^{\prime\prime}$. Applying Mason's gain formula, we readily obtain
\begin{align}
    G_{\UD}(x) = -\frac{\ps}{1-\ps} \left( 1 + \frac{\mathsf{a} x - 1}{1 - \mathsf{b} x + \mathsf{c} x^2}\right)
    \label{eq:genFun_Y}
\end{align}
where we introduced the ancillary quantities {$\mathsf{a}=1-\pBG$}, \mbox{$\mathsf{b}=(1-\pGB)(1-\ps) + (1 - \pBG)$}, and {$\mathsf{c}=(1-\ps)(1-\pGB - \pBG)$}.

The compact formulation in \eqref{eq:genFun_Y} is particularly useful, as it provides a complete characterization of the performance metrics under study for any configuration of system parameters,  in terms of both terminal activity and channel evolution. From this standpoint, indeed, closed form expressions for the average penalty function \avPenFunM\ can be readily derived from \eqref{eq:avPenFun_moments} for any value of $m$, by applying \eqref{eq:momentK_genFun}. For example, basic calculations lead to the average AoI
\begin{align}
    \avPenFun^{(1)} = \frac{1}{\ps} + \frac{\ps + \pGB(1-\ps)}{\pBG \ps} - \frac{1}{\pBG + \pGB} - \frac{1}{2}
\end{align}
which extends the well known formulation of the AoI of \ac{SA} in the absence of erasures ($1/\ps - 1/2$) \cite{Yates17:AoI_SA,Munari21_TCOM_AoI}, to a general Gilbert-Elliot channel setup.

The derived generating function allows also to obtain an expression for the complete \ac{PMF} of the r.v. \UD. As shown in \eqref{eq:genFun_Y}, in fact, $G_{\UD}(x)$ is given by the sum of a term of degree zero ($-\ps/(1-\ps)$) and of a proper rational function. Let us denote for convenience the latter as $A(x)/B(x)$. Applying partial fraction decomposition, this ratio can be expressed as
\begin{align}
    -\frac{\ps}{1-\ps} \cdot \frac{\mathsf{a} x - 1}{1 - \mathsf{b} x + \mathsf{c} x^2} = \frac{u_1}{\rho_1-x} + \frac{u_2}{\rho_2-x}
    \label{eq:partial_decomposition_interm}
\end{align}
where $\rho_1$ and $\rho_2$ are the roots of the degree-two denominator, while the combination coefficients $u_i$, $i=1,2$ evaluate to \mbox{$u_i = -A(\rho_i)/B^\prime(\rho_i)$}, denoting by $B^\prime(x)$ the first order derivative of $B(x)$. In turn, each of the two partial fractions can be conveniently expanded into a power series \cite[Ch.~11]{Feller}:\\[-.7em]
\begin{align}
    \frac{u_i}{\rho_i-x} = \frac{u_i}{\rho_i} \cdot \sum_{y=0}^{\infty} \left(\frac{x}{\rho_i} \right)^y .
\end{align}
Plugging these expressions into \eqref{eq:genFun_Y} we finally have 
\begin{align}
    G_\UD(x) = -\frac{\ps}{1-\ps} + \sum_{y=0}^{\infty} \left(\frac{u_1}{\rho_1^{y+1}} + \frac{u_2}{\rho_2^{y+1}}\right) x^y
\end{align}
which, recalling the definition of generating function given in \eqref{eq:genFun_def} offers the sought \ac{PMF}\footnote{Note that, by setting $x=0$ in \eqref{eq:partial_decomposition_interm}, we obtain $u_1/\rho_1 + u_2/\rho_2 = \ps/(1-\ps)$ for any parameters configuration, so that $P_\UD(0)=0$. This is expected, as at least one slot has to elapse between two successful updates.} 
\begin{align}
    P_\UD(\ud) = \frac{u_1}{\rho_1^{y+1}} + \frac{u_2}{\rho_2^{y+1}}\,, \quad y\geq 1.
    \label{eq:pmf_Y}
\end{align}

Leaning on this result, a compact closed form expression of the peak violation probability for any penalty order $m$ immediately follows \eqref{eq:peakPen_momnets}
\begin{align}
    \peakPenM(\theta) = \frac{u_1 \rho_1^{-\lfloor \theta^{1/m}\rfloor - 1} }{\rho_1-1} + \frac{u_2\rho_2^{-\lfloor \theta^{1/m}\rfloor - 1} }{\rho_2-1}.
    \label{eq:peakPen_final}
\end{align}

\begin{remark}To prove the ergodicity of $\penFunMRP(t)$, consider the embedded process $\penFunMRP_\ell$ and the discrete time \ac{MC} $(C_\ell,\penFunMRP_\ell)$ introduced at the beginning of this section. The infinite-state \ac{MC} can easily be shown to be irreducible and aperiodic. Observe that the mean recurrence time of state $(\good,0)$ \-- where \good\ denotes  good channel conditions \-- is $\expOp[\UD]$, which was shown to be strictly positive. Recalling that this is also the reciprocal of the stationary probability for the state, the existence of a proper distribution (and the ergodicity of the process) follows.
\end{remark}

\begin{remark}
Observe that the knowledge of the \ac{PMF} of the random variable $Y$ enables to derive the distribution of a broad class of functions of $Y$. This enables the analysis of age penalty function well beyond the power law adopted in this paper, by simply evaluating \eqref{eq:avPenFun_deriv_step1} accordingly.
\end{remark}
\section{Results and Discussion} \label{results}

\begin{figure}
    \centering
    \includegraphics[width=.8\columnwidth]{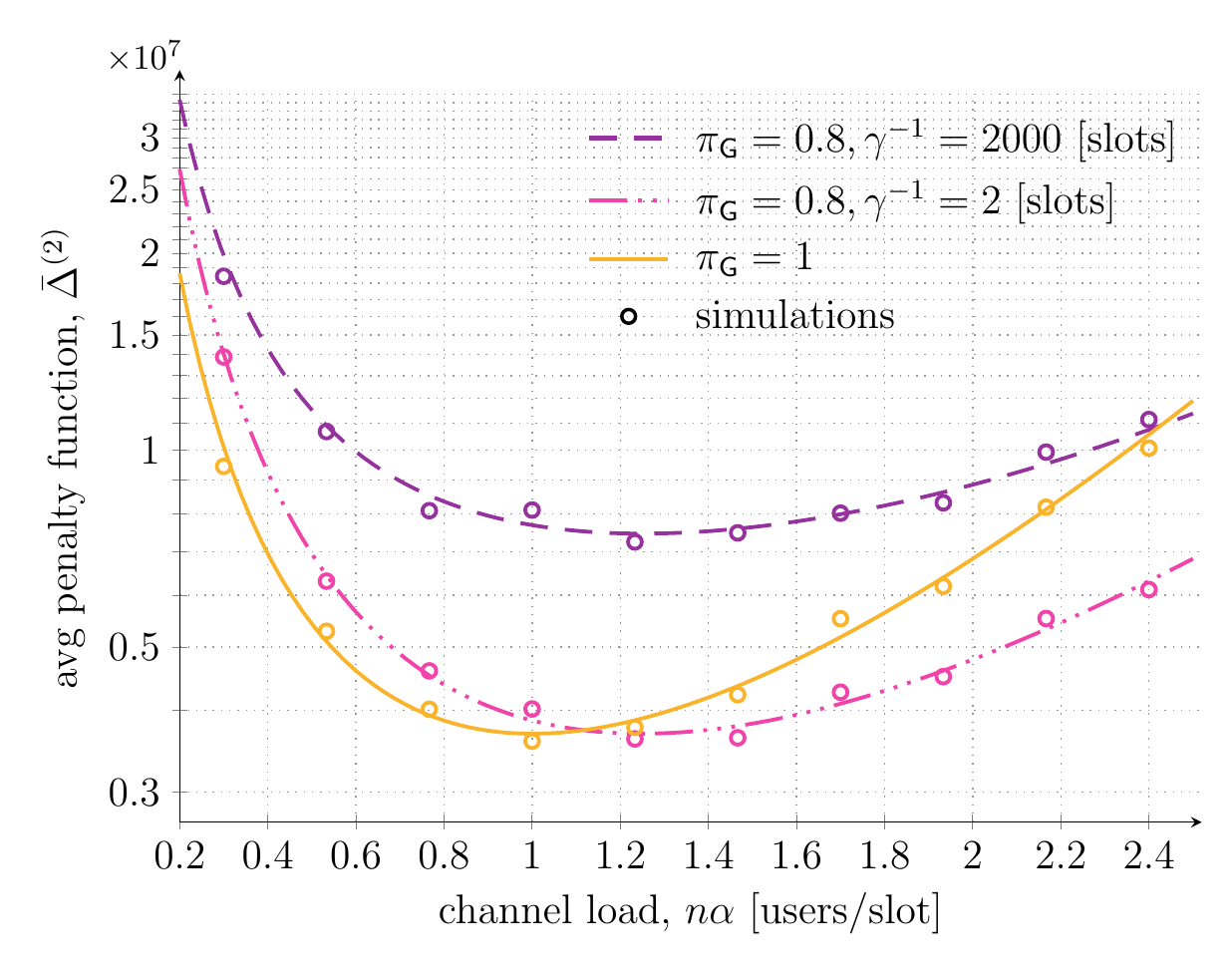}
    \vspace{-.7em}
    \caption{Average penalty function of order $m=2$ (\avPenFunTwo) vs. channel load ($\nodes\pGen$). Results obtained for $\nodes=500$. Markers denote simulation results.}
    \vspace{-4mm}
    \label{fig:avgPenalties}
\end{figure}

The framework developed in Sec.~\ref{sec:analysis} allows to study the behaviour of the considered \ac{SA} system under any configuration of channel parameters and traffic generation intensity \pGen, as well as for any order $m$ of the penalty function. To shed light on the key emerging tradeoffs we focus on a setup with $\nodes=500$ users, and start by studying the average penalty function for $m=2$. The metric is shown in Fig.~\ref{fig:avgPenalties} against the channel load $\nodes\pGen$, i.e. the average number of users that transmit an update over a slot. The solid line refers to operations over an ideal channel with no erasures ($\pBG=1$, $\pGB=0$, so that $\pGood=1$). In turn, non-solid lines are obtained for a channel which offers good conditions for a fraction $\pGood=0.8$ of the time. Specifically, we consider two insightful configurations. The former (dash-dotted line) is characterised by short bursts of erasures, with the average time spent by the channel in bad state prior to returning to a good state given by $1/\pBG = 2$ slots. The latter, instead, captures longer average periods of erasures, with $1/\pBG = 2000$ slots (dashed line).\footnote{We recall that the number of consecutive slots spent in bad state prior to moving back to a good one follows a geometric distribution of mean $\pBG^{-1}$. The two curves are obtained setting \pGB\ so to have an average time spent in good conditions $\pGood=0.8$.} In all settings, the well-known behaviour of \ac{SA} in terms of average information freshness metrics emerges \cite{Yates17:AoI_SA,Munari21_TCOM_AoI}. For low values of \pGen, \avPenFunTwo\ increases as the node generates updates too sporadically, whereas at higher channel loads performance degrades again as a result of network congestion (i.e., transmitted updates are lost due to collisions). The plot also highlights how average channel conditions determine the operating point that minimises the penalty function. Indeed, the optimal behaviour in terms of \avPenFunM\ is achieved by maximising the probability for a node to deliver an update over a slot, which, recalling \eqref{eq:psucc} is given by $\pGood\pGen(1-\pGood\pGen)^{\nodes-1}$. Regardless of the penalty order $m$, then, the channel load that minimises the average penalty readily evaluates to $1/\pGood$.

On the other hand, Fig.~\ref{fig:avgPenalties} insightfully reveals the critical role played by the duration of erasure bursts. This is well exemplified by the dashed and dash-dotted lines in the plot, which, albeit being characterised by the same average channel conditions ($\pGood=0.8$),  exhibit a remarkably different behaviour. In particular, an increase in the probability of undergoing longer periods of bad channel state has a detrimental effect on the average penalty function, regardless of the load. During such periods, in fact, no update can be delivered by the node, leading to a significant growth in the penalty undergone at the receiver. Notably, such an effect more than offsets the concurrent increase in the average duration of bursts of good channel conditions, resulting in worse information freshness. 

\begin{figure}
    \centering
    \includegraphics[width=.8\columnwidth]{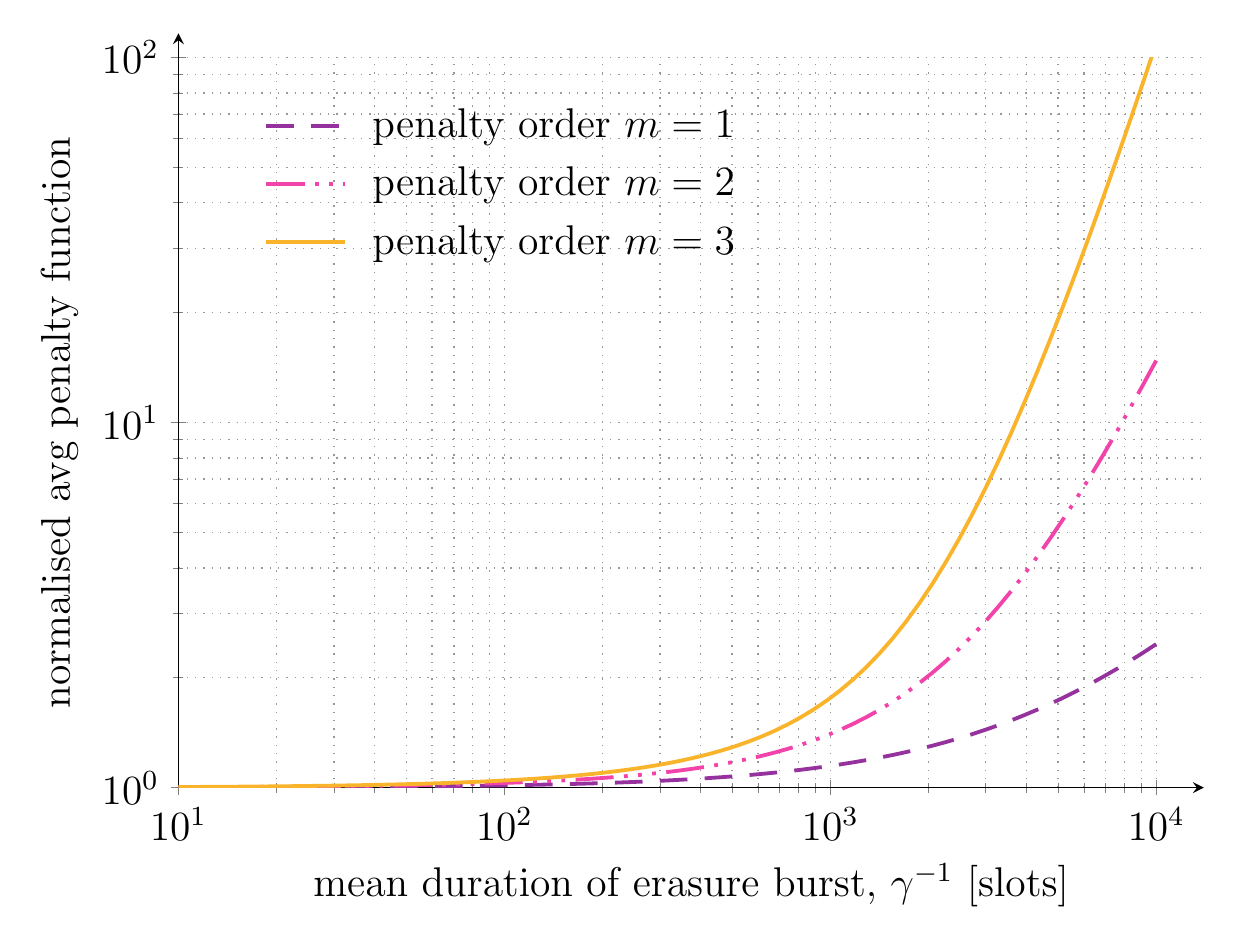}
    \vspace{-.7em}
    \caption{Ratio of \avPenFunM\ obtained for a given value of $\pBG$ to the same function obtained for $\pBG=1$. Different lines report the behaviour for distinct penalty orders $m$. In all cases, $n=500$, $\pGood = 0.8$, and $\nodes\pGen=1/\pGood$.}
    \label{fig:ratios}
\end{figure}

This aspect is further explored in Fig.~\ref{fig:ratios}, where we consider \mbox{$\pGood=0.8$} and a system operated at the optimal channel load $\nodes\pGen = 1/\pGood$. For different penalty orders $m$, the plot shows the ratio of the value of \avPenFunM\ obtained for a certain average duration of an erasure burst ($\pBG^{-1}$) to the value of the same penalty function  achieved for $\pBG=1$ (i.e., for the minimum possible burst duration). The plot offers additional insights, clarifying how the detrimental effect of longer periods of erasures becomes more pronounced as $m$ increases, with the least impact experienced in terms of average AoI ($m=1$). 
\begin{figure}
    \centering
    \includegraphics[width=.8\columnwidth]{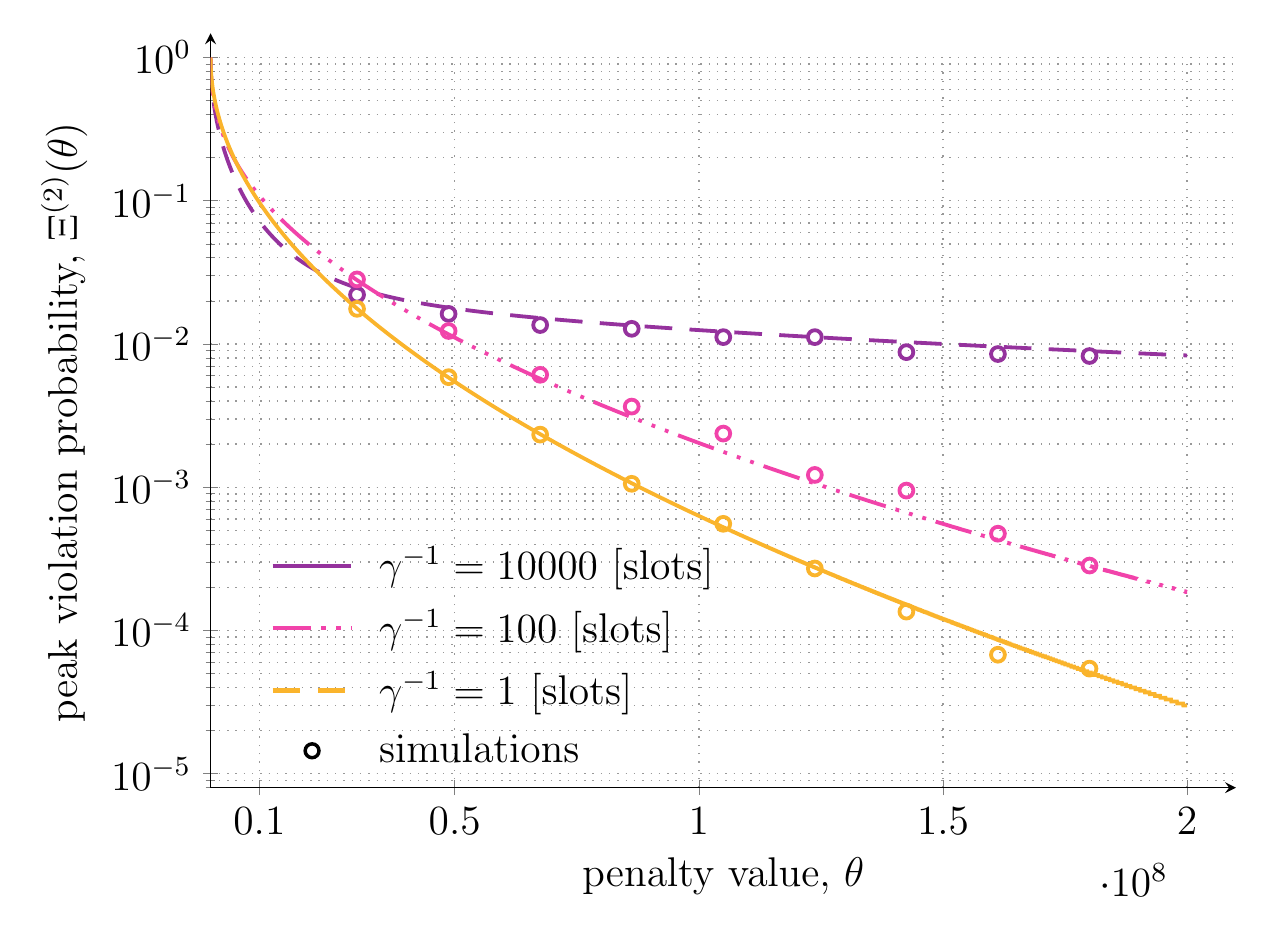}
    \vspace{-.7em}
    \caption{Peak violation probability for penalty order $m=2$. Results for $n=500$, $\pGood = 0.8$, and channel load $\nodes\pGen=1/\pGood$. Markers denote simulations.}\vspace{-3mm}
    \label{fig:ccdf}
\end{figure}

The system behaviour in terms of peak violation probability is reported in Fig.~\ref{fig:ccdf}, which shows $\peakPen^{(2)}$ assuming $\pGood=0.8$ and a  load $\nodes\pGen=1/\pGood$. The impact of channel conditions is once more apparent, with distinct values of $\pBG$ leading to significantly different distribution tails. From this standpoint, the analysis clarifies how channels with shorter bursts of errors tend to favour the maintenance of fresh information at the receiver. In this perspective, the design of advanced update generation and delivery strategies emerges as a fundamental step forward, calling for the extension of early research results \cite{Uysal21_AlohaThresh} to more complex channel models and to information freshness metrics that go beyond \ac{AoI}. In conclusion, we observe that the presented framework also provides simple yet powerful design tools in a broad range of settings, as highlighted in the following example.

\begin{figure}
    \centering
    \includegraphics[width=.8\columnwidth]{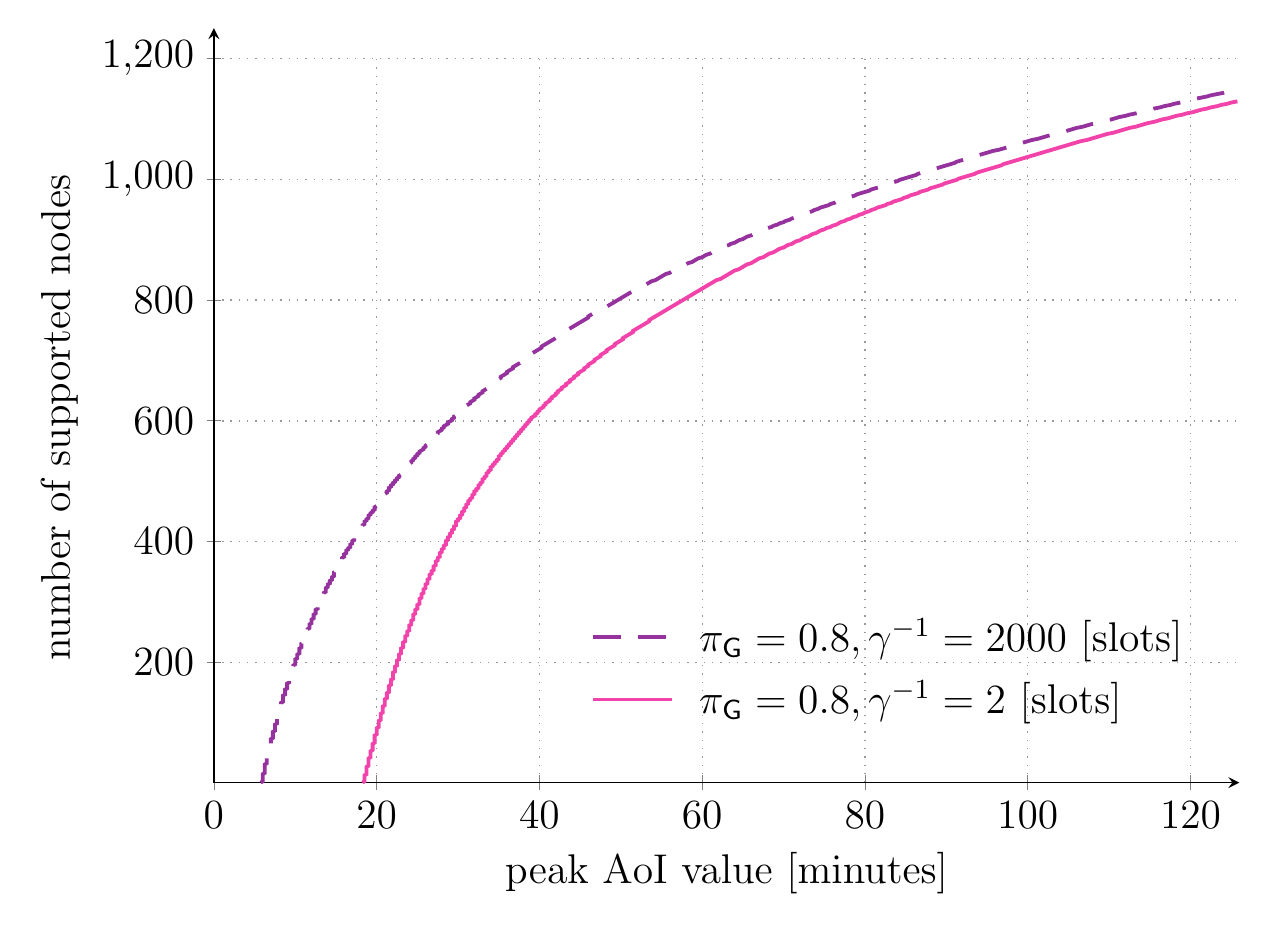}
    \vspace{-.7em}
    \caption{Number of supported nodes to not exceed a maximum peak AoI value with probability lower than $10^{-3}$. Parameters set as for Example 1.}\vspace{-3mm}
    \label{fig:numNodes}
\end{figure}

\begin{example}Consider an \ac{IoT} system in which nodes share a \ac{SA} channel to report sensed data to a gateway. Taking typical LoRaWan standard parameters, assume that time slots have duration of $136$ ms (due to $32$ bytes payload and transmission bandwidth of $125$ kHz) and that each node generates an update on average every $5$ minutes. In this setting, an operational constraint common to many  applications is to keep the probability of exceeding a maximum level of a penalty function below a desired threshold. Assume for instance that \ac{AoI} (i.e., $m=1$) is of interest. Fig.~\ref{fig:numNodes} reports the number of nodes that can be supported not to exceed with probability lower than $10^{-3}$ the peak \ac{AoI} value reported on the $x$-axis, obtained from \eqref{eq:peakPen_final}.
\end{example}

\bibliographystyle{IEEEtran}
\bibliography{IEEEabrv,biblio_AoI}

\begin{thebibliography}{10}
\providecommand{\url}[1]{#1}
\csname url@samestyle\endcsname
\providecommand{\newblock}{\relax}
\providecommand{\bibinfo}[2]{#2}
\providecommand{\BIBentrySTDinterwordspacing}{\spaceskip=0pt\relax}
\providecommand{\BIBentryALTinterwordstretchfactor}{4}
\providecommand{\BIBentryALTinterwordspacing}{\spaceskip=\fontdimen2\font plus
\BIBentryALTinterwordstretchfactor\fontdimen3\font minus
  \fontdimen4\font\relax}
\providecommand{\BIBforeignlanguage}[2]{{%
\expandafter\ifx\csname l@#1\endcsname\relax
\typeout{** WARNING: IEEEtran.bst: No hyphenation pattern has been}%
\typeout{** loaded for the language `#1'. Using the pattern for}%
\typeout{** the default language instead.}%
\else
\language=\csname l@#1\endcsname
\fi
#2}}
\providecommand{\BIBdecl}{\relax}
\BIBdecl

\bibitem{Yates20_Survey}
R.~D. Yates, Y.~Sun, D.~R. Brown, S.~K. Kaul, E.~Modiano, and S.~Ulukus, ``Age
  of information: An introduction and survey,'' \emph{{IEEE} J. Sel. Areas
  Commun.}, vol.~39, no.~5, pp. 1183--1210, May 2021.

\bibitem{Yates17:AoI_SA}
R.~Yates and S.~K. Kaul, ``Status updates over unreliable multiaccess
  channels,'' in \emph{Proc. IEEE ISIT}, 2017.

\bibitem{Munari21_TCOM_AoI}
A.~Munari, ``Modern random access: an age of information perspective on
  irregular repetition slotted {ALOHA},'' \emph{{IEEE} Trans. Commun.},
  vol.~69, no.~6, pp. 3572--3585, Jun. 2021.

\bibitem{Uysal21_AlohaThresh}
O.~T. Yavaskan and E.~Uysal, ``Analysis of slotted {ALOHA} with an age
  threshold,'' \emph{{IEEE} J. Sel. Areas Commun.}, vol.~39, no.~5, pp.
  1456--1470, May 2021.

\bibitem{Ephremides20_AoII}
\BIBentryALTinterwordspacing
A.~Maatouk, M.~Assaad, and A.~Ephremides, ``The age of incorrect information:
  an enebler of semantic-empowered communication,'' 2020. [Online]. Available:
  \url{http://arxiv.org/abs/2012.13214v1}
\BIBentrySTDinterwordspacing

\bibitem{Uysal20_TIT}
Y.~Sun, Y.~Polyanskiy, and E.~Uysal, ``Sampling of the {Wiener} process for
  remote estimation over a channel with random delay,'' \emph{{IEEE} Trans.
  Inf. Theory}, vol.~66, no.~2, pp. 1118--1135, Feb. 2020.

\bibitem{Gilbert60}
E.~{Gilbert}, ``Capacity of a burst-noise channel,'' \emph{The {Bell} System
  Technical Journal}, vol.~39, no.~5, pp. 1253--1265, 1960.

\bibitem{Elliott63}
E.~O. Elliott, ``Estimates of error rates for codes on burst-noise channels,''
  \emph{Bell Syst. Tech. J.}, vol.~42, pp. 1977--1997, Sep. 1963.

\bibitem{turin1999digital}
W.~Turin, \emph{Digital transmission systems: performance analysis and
  modeling}.\hskip 1em plus 0.5em minus 0.4em\relax McGraw-Hill Companies,
  1999.

\bibitem{Zorzi:GE}
M.~Zorzi, R.~R. Rao, and L.~B. Milstein, ``Error statistics in data
  transmission over fading channels,'' \emph{{IEEE} Trans. Commun.}, vol.~46,
  no.~11, pp. 1468--1477, Nov. 1998.

\bibitem{lutz2012satellite}
E.~Lutz, M.~Werner, and A.~Jahn, \emph{Satellite systems for personal and
  broadband communications}.\hskip 1em plus 0.5em minus 0.4em\relax Springer
  Science \& Business Media, 2012.

\bibitem{shannon1942theory}
C.~E. Shannon, ``The theory and design of linear differential equation
  machines,'' Princeton University, Princeton, N. J., Nat'l Defense Res. Com.
  Rept., Jan. 1942.

\bibitem{Abramson77:PacketBroadcasting}
N.~Abramson, ``{The throughput of packet broadcasting channels},'' \emph{{IEEE}
  Trans. Commun.}, vol. COM-25, no.~1, pp. 117--128, Jan. 1977.

\bibitem{Chiariotti21_peakAge}
F.~Chiariotti, O.~Vikhrova, B.~Soret, and P.~Popovski, ``Peak age of
  information distribution for edge computing with wireless links,''
  \emph{{IEEE} Trans. Commun.}, vol.~69, no.~5, pp. 3176--3191, May 2021.

\bibitem{Durisi19_JSAC}
R.~{Devassy}, G.~{Durisi}, G.~C. {Ferrante}, O.~{Simeone}, and E.~{Uysal},
  ``Reliable transmission of short packets through queues and noisy channels
  under latency and peak-age violation guarantees,'' \emph{{IEEE} J. Sel. Areas
  Commun.}, vol.~37, no.~4, pp. 721--734, Apr. 2019.

\bibitem{NLAoI}
A.~Kosta, N.~Pappas, A.~Ephremides, and V.~Angelakis, ``Age and value of
  information: {N}on-linear age case,'' in \emph{Proc. IEEE ISIT}, 2017.

\bibitem{Telatar18}
E.~{Najm} and E.~{Telatar}, ``Status updates in a multi-stream {M/G/1/1}
  preemptive queue,'' in \emph{{Proc. IEEE INFOCOM}}, 2018.

\bibitem{Lorens56}
C.~Lorens, ``Theory and applications of signal flow graphs,'' \emph{{MIT Tech.
  Report}}, 1956.

\bibitem{Mason55}
S.~Mason, ``Feedback theory \--- further properties of signal flow graphs,''
  \emph{Proceedings of the IRE}, vol.~44, no.~7, pp. 920--926, Jul. 1955.

\bibitem{Feller}
W.~Feller, \emph{An introduction to probability theory and its applications,
  Vol. 1}.\hskip 1em plus 0.5em minus 0.4em\relax {New York}: Wiley, 1957.

\end{thebibliography}

\end{document}